# Photon-energy-programmable subnanometric electron birth-site control

**Hirofumi Yanagisawa[1]✉, Abhisek Sinha[1], Ravi Kumar[2], Hiroki Funata[2], Pasquale Barbato[1], Neill Lambert[3], Hirotaka Kitoh-Nishioka[2]**

[1] Research Institute of Electronics (RIE), Shizuoka University, Hamamatsu, Japan

[2] Department of Energy and Materials Engineering, Faculty of Science and Engineering, Kindai University, Higashi-Osaka, Japan

[3] RIKEN Center for Quantum Computing (RQC), Wako, Saitama, Japan

✉ Correspondence to: yanagisawa.hirofumi@shizuoka.ac.jp

Optical control of electron-generation sites has broadly enabled ultrafast nanoscale imaging, spectroscopy, and functional control. Existing approaches achieve nanoscale site selectivity by shaping localised optical fields around nanostructures, thereby limiting independent site selectivity within the same local-field hotspot. Here, using a single-molecule electron emitter, we show that site selectivity can instead be encoded in the electronic excitation pathway, enabling subnanometric control of electron birth sites within a common local-field hotspot. By tuning the photon energy, we selectively access different groups of molecular states whose combined contributions define emission channels with distinct spatial character, thereby reversibly switching the electron birth site within the same emitter. The switching is read out directly in the far-field emission pattern and occurs upon changing the photon energy, but not upon varying the intensity or polarisation. Our results establish optical birth-site selectivity that is not dictated by the local-field distribution, opening a route to electron birth-site control through the electronic excitation pathway.

The ability to optically define where electrons are generated is central to ultrafast nanoscale imaging[1–3], spectroscopy[4–6], and functional control[7–9]. In established approaches, such site selectivity is achieved by shaping the real-space distribution of localised optical fields through nanostructure design and optical control parameters such as polarisation, phase, and amplitude[7–9]. For example, as illustrated schematically in Fig. 1a, varying these parameters shifts the optical hotspot across a triangular nanostructure, allowing emission to be switched between sites such as A and B[2,3]. In principle, shrinking the nanostructure can further reduce the hotspot size, and atomic-scale structural features can confine the field to very small length scales[10–12], suggesting that selection on the scale of a few nanometres may be possible[12]. Yet the selected site is still dictated by the real-space local-field profile, and if the field distribution is left unchanged, there is no independent optical handle for selecting substructures within the same local-field hotspot.

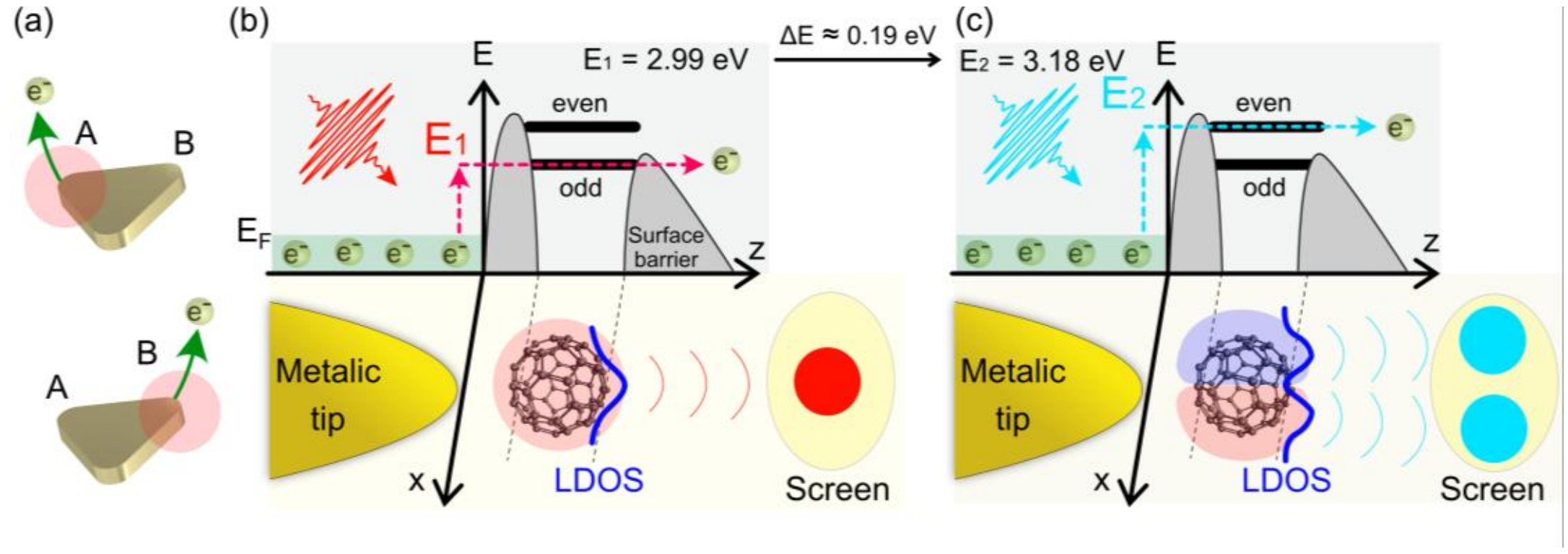

**Fig. 1 | Concept of photon-energy-dependent subnanometric control of electron-emission sites.**

**a**, Schematic of conventional field-based optical control of electron-emission sites. Varying optical parameters shifts the localised optical-field hotspot (red shaded area) across the triangular nanostructure, switching the dominant emission site between positions such as A and B. In this scheme, site selectivity is determined by the real-space position of the local-field hotspot. **b**,**c**, Schematic diagrams of a simplified single-molecule electron emitter and the corresponding potential profiles. Two closely spaced excitation energies, $E_1$ and $E_2$, selectively couple to molecular orbitals with distinct spatial symmetries, leading to different dominant emission channels. A small change in excitation energy thus enables reversible relocation of the emission site within a single molecule. Electronic states in the metal are occupied up to the Fermi level, $E_F$, as indicated schematically in green. The grey profile schematically indicates the surface potential barrier through which photoexcited electrons tunnel into vacuum under the applied DC field. The corresponding near-field emission distributions then evolve into distinct far-field patterns under strong-field acceleration. The thick blue curves schematically denote the LDOS associated with the contributing states.

As a result, site selectivity remains fundamentally limited whenever multiple substructures share a common hotspot.

By contrast, here we show that site selectivity can be encoded in the electronic excitation pathway, enabling subnanometric control within an essentially unchanged local-field hotspot. In this scheme, the key control parameter is not where the optical field is localised, but which electronic states are selectively accessed by photoexcitation. To explore this concept, we use a single-molecule electron emitter formed by a molecular protrusion at the apex of a metallic tip, as schematically illustrated in Fig. 1b,c[13]. A strong DC field is applied to the tip, establishing conditions under which electrons photoexcited from occupied states near the Fermi level can tunnel through the surface potential barrier into vacuum, as illustrated by the accompanying potential diagram. Previous work showed that the emission site in such an emitter is highly sensitive to the molecular orbital through which the electron is emitted[14]. In that work, subnanometric modulation of the effective emission region was achieved by switching between DC field emission and laser-induced photoemission, which involved different molecular-orbital-mediated emission channels; however, it did not demonstrate switching between two optically driven emission conditions, as subsequently discussed[15]. Building on this idea, we reasoned that photon energy alone could be used to switch the emission site by selectively accessing molecular states with distinct spatial symmetries. As illustrated in Fig. 1b,c, we use two nearby photon energies, $E_1$ and $E_2$, to excite the same emitter. Although a single molecule contains many discrete states, several of which can contribute within the finite energy range of the photoexcited electrons, the schematic is deliberately simplified by representing their collective contribution at each photon energy with a single idealized molecular state. In Figs. 1b and 1c, these representative states are depicted as having predominantly odd-like and even-like spatial character, respectively, to illustrate how a photon-energy-dependent redistribution among the contributing states can produce emission channels with distinct spatial characteristics. Previous theoretical work suggests that the emission site reflects the local density of states (LDOS) of the contributing states[16]. The associated LDOS is indicated schematically by the thick blue curves. The different spatial symmetries accessed at $E_1$ and $E_2$ therefore shift the effective electron birth site within the same molecule. The resulting change can then be read out directly from the far-field electron distribution on the detector screen.

Here we realise this concept experimentally in a single-molecule electron emitter. By tuning the excitation energy within a narrow range, we selectively access different groups of molecular states whose combined contributions define emission channels with distinct spatial character, thereby reversibly switching the electron birth site within the same emitter. The switching depends exclusively on photon energy and is absent under variations in optical intensity or polarisation.

These results demonstrate photon-energy-controlled switching of the electron birth site on the subnanometric scale within a common optical hotspot, thereby overcoming a key limitation of conventional local-field patterning.

**Reversible photon-energy-dependent switching in a single-molecule emitter**

Electron emission patterns were recorded from single-molecule emitters under ultrahigh vacuum. The emitters were formed stochastically at the apex of a tungsten tip after deposition of $C_{60}$ multilayers and removal of most molecules under a strong DC field[13]. Measurements were performed on the same emitter using femtosecond laser pulses with photon energies of either 2.99 or 3.18 eV, while excitation intensity and polarisation were independently controlled. The higher photon energies used here (~3 eV), compared with our previous ~1.5 eV experiments[14], enabled operation at substantially lower optical powers (typically ~1 mW), reducing thermal load and improving emitter stability so that measurements could be sustained for about 1 h before molecular desorption. The average number of detected electrons remained below 0.05 per pulse, so that multi-electron emission and space-charge effects were negligible.

The imaging geometry used here is based on the principle of field emission microscopy, in which electrons emitted from the tip apex are accelerated toward the detector and the resulting detector pattern retains a spatial correlation with the electron-emission distribution at the tip surface[9]. For the single-molecule emitters studied here, the effective spatial resolution of the emission patterns, when referred back to the emitter, has been theoretically estimated to be ~0.3 nm[17], corresponding to an intramolecular length scale. This estimate is consistent with our previous experimental observations[14]. The measured patterns therefore provide spatially resolved experimental information about changes in the emission distribution. The corresponding intramolecular length scale is estimated by relating the measured far-field patterns to the simulated near- and far-field distributions, as discussed below. Experimental details are provided in the Methods section, with additional information in Sections S1 and S2 of the Supplementary Information.

Fig. 2 shows the photon-energy-dependent switching of electron emission patterns on the detector screen. The upper panels of Fig. 2a show representative laser-induced far-field emission patterns recorded from the same single-molecule electron emitter at excitation energies $E_1$ and $E_2$. At $E_1$, the pattern exhibits an oval ring-like structure, whereas at $E_2$ it becomes distinctly two-lobed. The pronounced difference between the two patterns is further highlighted by the contour overlays in Fig. 2a. Fig. 2b shows that this switching is reproducible and fully reversible within the same emitter upon cycling the excitation energy between $E_1$ and $E_2$, demonstrating relocation of the emission channel without irreversible modification of the emitter. Photon-energy-dependent changes in the emission pattern are also observed for multiple emitters (Fig. 2c and Fig. S2), although the particular combination of patterns observed at $E_1$ and $E_2$ can vary among emitters. For example, Mol. 2 shows a change in the separation between the two lobes, whereas Mol. 3 shows a transformation from a two-lobed pattern to a more connected structure. These observations indicate that photon-energy-dependent control of the emission pattern is not restricted to a particular local arrangement at the emitter apex. The experimentally observed patterns are also reproduced by simulations (Fig. 2a, lower panels), consistent with the interpretation that the switching arises from a photon-energy-dependent redistribution among molecular emission channels with distinct spatial character, as discussed below. Because $E_1$ and $E_2$ are closely spaced in photon energy, any accompanying change in the optical-field distribution is expected to be minor. The observed switching therefore most naturally reflects a change in the emission channel within essentially the same optical hotspot.

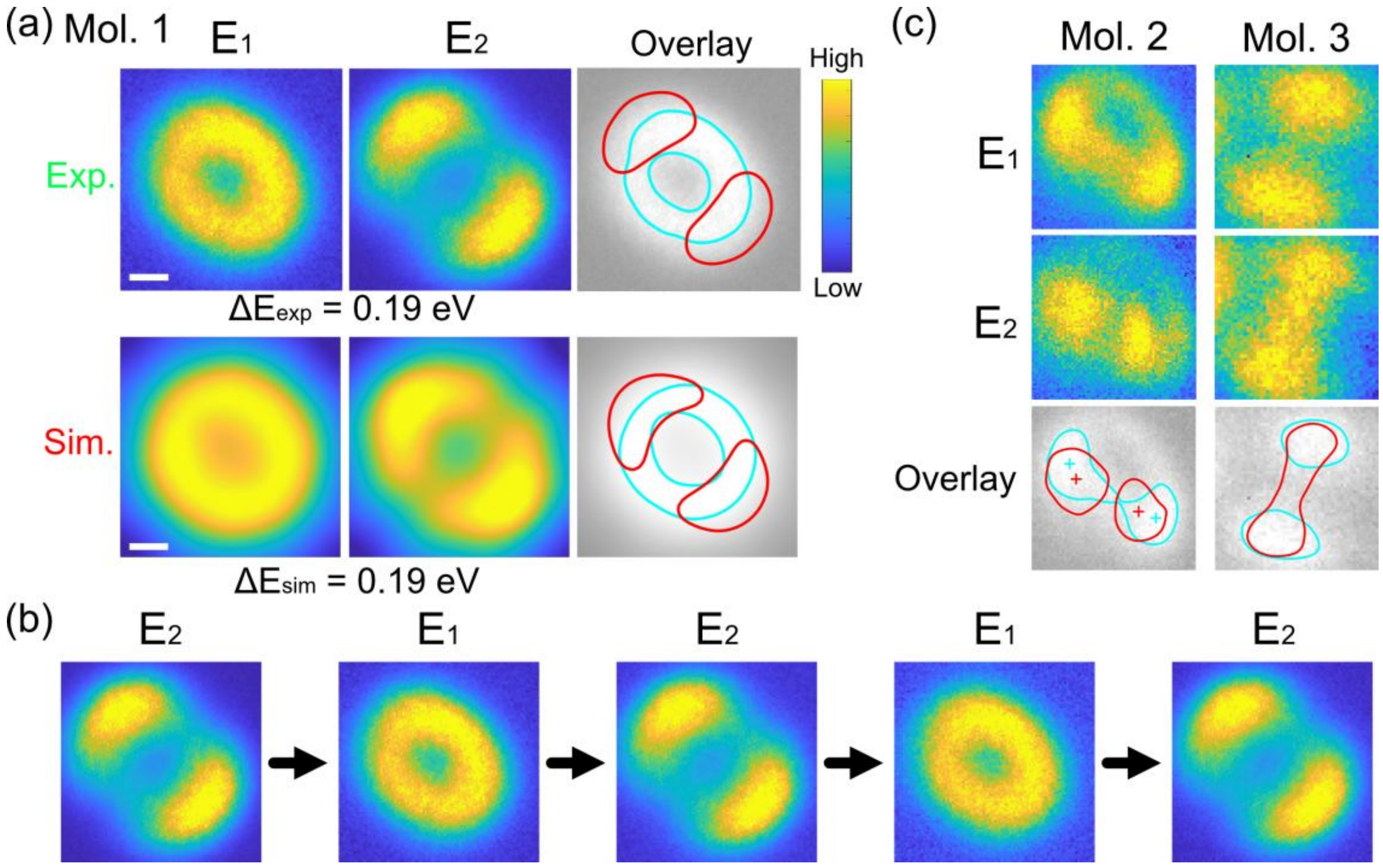

Fig. 2 | Reversible photon-energy-dependent switching of molecular emission patterns. **a**, Electron emission patterns on the detector screen recorded from the same single-molecule emitter at excitation energies $E_1$ and $E_2$. Distinct patterns are observed, consistent with selective activation of different molecular emission channels by photon energy. Experimental contour overlays are shown at right. The corresponding simulated patterns are shown below, with their contour overlays at far right. White bars indicate scale bars. Scale bars, 3 mm (experiment and simulation). **b**, Reversible switching between the two emission patterns from the same molecule. Alternating the excitation energy between $E_1$ and $E_2$ repeatedly switches the emission pattern, demonstrating that the process is reproducible and does not result from irreversible structural changes in the emitter. **c**, Additional examples of single-molecule emitters exhibiting photon-energy-dependent changes in their emission patterns. The specific patterns and their changes differ among emitters. In the overlay for Mol. 2, crosses indicate the centres of mass of the emission lobes.

## Switching is governed by excitation energy, not by local-field redistribution

To distinguish the switching observed in Fig. 2 from the conventional local-field patterning mechanism, we next examine its dependence on polarisation and excitation intensity. At fixed excitation energy, rotating the polarisation leaves the emission pattern unchanged (Fig. 3a), as further highlighted by the overlap of the contour lines. To quantify this behaviour, we define $d$ as the separation between the two emission lobes and $d_0$ as the corresponding value in the leftmost image of each measurement series, which is used as the reference pattern. The normalised lobe separation, $d/d_0$, remains constant within experimental uncertainty, confirming that polarisation rotation does not alter the spatial configuration of the emission pattern. Likewise, varying the excitation power changes the electron yield but does not modify the overall spatial configuration (Fig. 3b). The somewhat larger deviation of $d/d_0$ at low power is attributed to the reduced signal-to-noise ratio in this regime rather than to a genuine change in the emission geometry, as indicated by the substantial overlap of the contour lines. The electron yield increases linearly with excitation power (Fig. 3c), consistent with a single-photon excitation process and with the absence of any intensity-driven switching. These results distinguish the present mechanism from conventional local-field patterning approaches, in which the selected emission site is controlled through polarisation- or field-dependent redistribution of the optical near field. Instead, the switching observed here is governed exclusively by excitation energy.

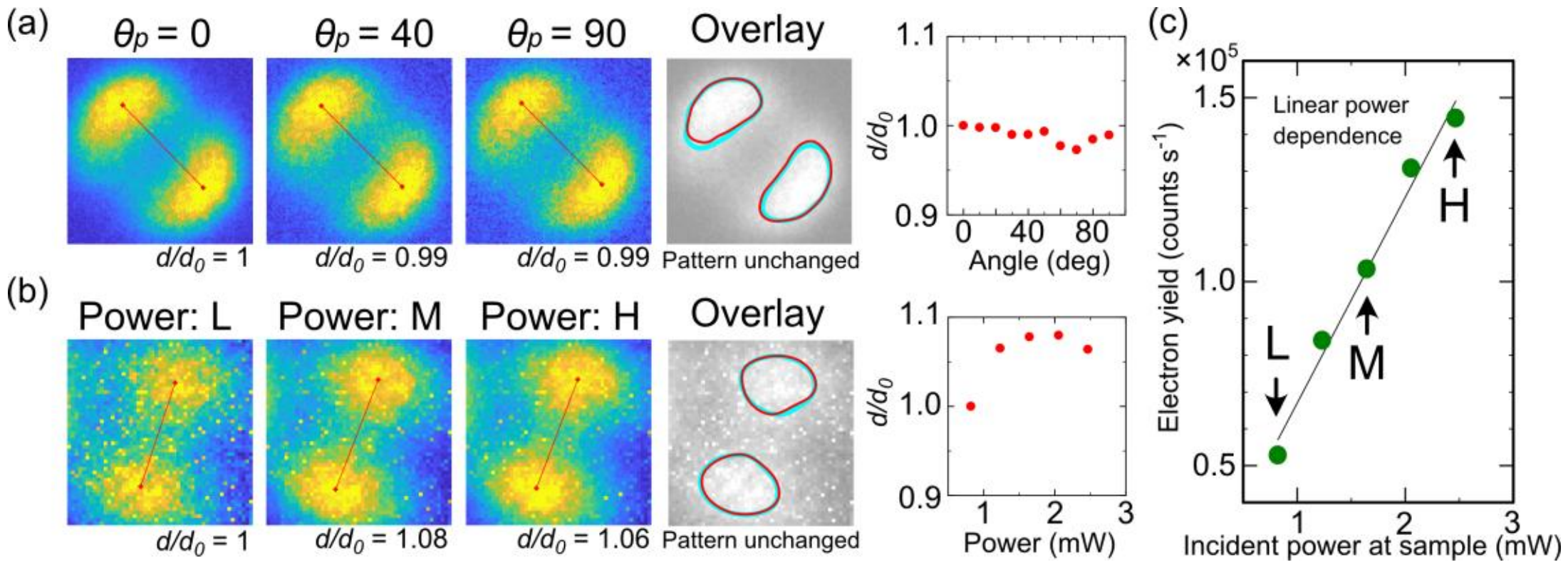

**Fig. 3 | Emission-pattern invariance under polarisation and power variations.**
**a**, Emission patterns of Mol. 1 shown in Fig. 2a recorded at fixed excitation energy $E_1$ while rotating the incident polarisation angle $\theta_p$ (0°, 40° and 90°). The spatial configuration remains unchanged. The polarisation angle $\theta_p$ is defined with respect to the tip axis; $\theta_p = 0°$ corresponds to polarisation parallel to the tip axis. **b**, Emission patterns of Mol. 2 in Fig. 2c recorded at fixed excitation energy $E_1$ for increasing incident power [low (L), medium (M) and high (H)]. Although the electron yield increases, the spatial configuration remains unchanged. In **a** and **b**, images are displayed on a common linear intensity scale after identical percentile clipping. The spatial configuration is quantified by the normalised lobe separation $d/d_0$, where $d$ is the distance between the lobe centres of mass and $d_0$ is the corresponding value for the reference image in each series ($\theta_p = 0$ in **a**; the lowest-power image in **b**). The corresponding $d/d_0$ values are plotted at right. No systematic variation is observed with polarisation angle or incident power. **c**, Electron yield as a function of incident power at the sample, showing linear scaling consistent with a single-photon excitation process.

## Photon-energy-selective orbital access and subnanometric relocation of the emission channel

To interpret the switching mechanism more realistically than in the simplified picture of Fig. 1b,c, we model the observed emission pattern as an incoherent sum of contributions from multiple molecular orbitals. As schematically illustrated in Fig. 4a, optical excitation first produces a finite-width non-equilibrium electron distribution in the metallic tip. Energy-dependent transmission through the potential barriers reshapes this distribution into a relatively narrow, asymmetric post-barrier electron energy distribution[6], hereafter referred to as the energy-supply function. The magenta and cyan curves represent these distributions for $E_1$ and $E_2$, respectively. The change in photon energy shifts the peak position of the energy-supply function (see Section S5 of the Supplementary Information), thereby changing the group of molecular orbitals that is energetically accessible. Only orbitals lying within this energy window contribute appreciably to the emission. For each such orbital, an effective emission weight is determined by the overlap between the molecular orbital and the evanescent continuation of the photoexcited metal-state wavefunction into the molecular region, multiplied by the value of the energy-supply function at the corresponding orbital energy. The molecular orbitals were obtained from electronic-structure calculations, and the far-field emission pattern associated with each orbital was calculated by propagating the emission-relevant orbital component in a uniform electric field. The simulated detector pattern is constructed as the incoherent weighted sum of the orbital-resolved far-field intensities.

Figure 4b resolves the calculated emission into the effective emission weight and final far-field contribution of each relevant molecular orbital. Here, the effective emission weight characterises the relative supply of photoexcited electrons from the metal into each molecular orbital. The final far-field contribution includes this supply weight and additionally accounts for orbital-dependent coupling from the molecule into vacuum and subsequent propagation to the detector. At both $E_1$ and $E_2$, several orbitals within the finite-width energy-supply window contribute appreciably; neither emission channel can therefore be assigned exclusively to a single orbital. Instead, the

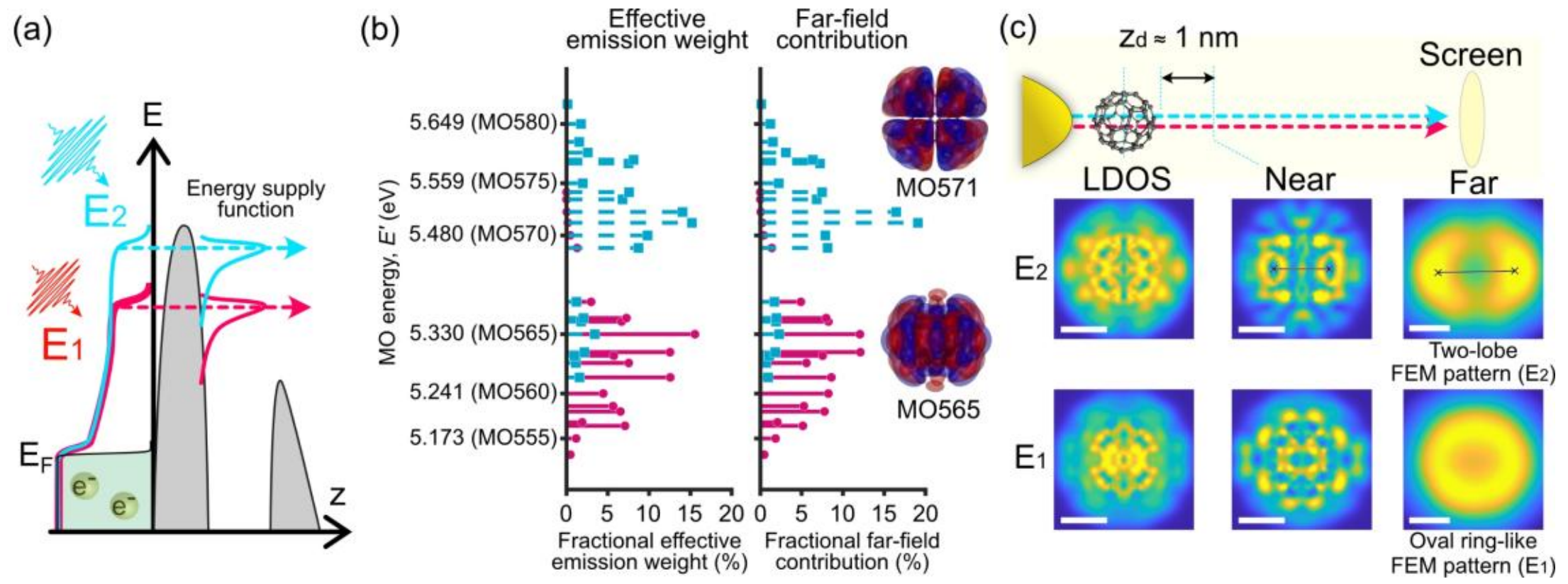

**Fig. 4 | Photon-energy-selected emission channels from the molecular scale to the far field.**

**a**, Schematic potential diagram of a single-molecule electron emitter. The green-shaded region represents the occupied electronic states in tungsten up to the Fermi level. The magenta and cyan curves represent the non-equilibrium electron distributions in the metal and the corresponding post-barrier electron energy distributions (energy-supply functions) for $E_1$ and $E_2$, respectively. **b**, Orbital-resolved effective emission weights (left) and far-field contributions (right) for $E_1$ (magenta) and $E_2$ (cyan). The effective emission weight combines the finite-width energy-supply function with the three-dimensional overlap between the photoexcited metal-state wavefunction and each MO. The far-field contribution further accounts for MO-dependent coupling from the molecule into vacuum and subsequent propagation towards the detector, and therefore represents the fractional contribution of each MO to the final simulated detector pattern. For each photon energy, the effective emission weights and far-field contributions are separately normalized to their respective totals. The vertical axis shows the MO energies on the energy coordinate used in the emission simulation, with selected MO indices indicated in parentheses. Representative wavefunction isosurfaces of MOs 565 and 571 are shown as insets, with red and blue indicating opposite signs of the wavefunction. **c**, LDOS, near-field, and far-field distributions for the dominant $E_1$ and $E_2$ channels. Near-field maps are evaluated at a distance $z_d$ from the molecular surface, where $z_d \approx 1$ nm. Crosses in the $E_2$ near- and far-field maps indicate the centres of mass used to quantify the dominant lobe separation. Scale bars, 0.5 nm (LDOS and near field) and 5 mm (far field). The orientation is chosen for clarity; the absolute azimuth is arbitrary.

shift of the energy-supply function redistributes the contributions among different groups of orbitals. At $E_1$, MOs 564 and 565 make comparably large contributions to the final far-field pattern, whereas MOs 571 and 572 contribute prominently at $E_2$. Other nearby orbitals also make non-negligible contributions. Because the resulting patterns arise from the complete weighted ensembles rather than from individual orbitals, we next examine how the spatial distributions of these collectively formed emission channels evolve from the molecular scale to the detector.

To follow this evolution, Fig. 4c compares the LDOS, near-field distribution, and far-field pattern associated with the $E_1$ and $E_2$ emission channels. The near-field maps are evaluated using the outward-emitting components of the relevant orbitals at a distance $z_d$ from the molecular surface, where $z_d \approx 1$ nm. The far-field maps are calculated separately by propagating the emission-relevant orbital components in a uniform electric field. In both cases, the maps are constructed as incoherent sums using the weighting described above. These comparisons trace how the distinct spatial character of each collectively formed emission channel develops from the molecular region through the near field into the final detector pattern.

Previous theoretical work showed that the spectroscopic weight of emission is governed by the LDOS of the contributing states, which in turn suggests that the emission site reflects their LDOS[16]. In our earlier experiments, this tendency was indeed observed in the far field within the spatial-resolution limit of the microscopy. In the present case, however, the correspondence between the LDOS and the far-field intensity is less direct, as seen in Fig. 4c. This difference appears to reflect the spatial character of the contributing molecular orbitals. In those earlier cases, emission was dominated by more superatomic-like orbitals that are spatially delocalised away from the molecular surface, with relatively weak atomic-scale modulation, and therefore showed a more direct correspondence between the LDOS and the emitted intensity[14,15]. In the present case, however, the dominant contributing orbitals retain a coarse-grained superatomic-like symmetry but exhibit more pronounced atomic-scale modulation, as illustrated by the representative orbitals shown in Fig. 4b (MOs 565 and 571). At higher energies, atomically modulated MOs are generally more spatially extended and therefore couple more strongly to the tunnelling channel, as discussed in Section S10 of the Supplementary Information.

To understand why the LDOS and the far-field intensity correspond less directly in the present case, we consider how the finer spatial structure of these orbitals is mapped onto the detector. A key point is that the atomic-scale modulation of the LDOS is distributed along the curved carbon framework of $C_{60}$, whereas the experimentally observed far-field pattern is formed on the detector plane after propagation. This curvature-to-plane mapping alters the apparent distribution during propagation. Accordingly, comparison between the near-field and far-field distributions is more informative here, because the near-field maps are evaluated on a plane and are therefore more directly comparable to the detector-plane distribution. Although atomically resolved features are not clearly visible in the far field, partly because of the finite spatial resolution of the microscopy, the coarse-grained symmetry of the emission channel is preserved.

Using the simulated near- and far-field patterns, we estimated the characteristic real-space scale associated with the two-lobe far-field pattern. The centres of mass of the two dominant near-field lobes, marked by crosses in Fig. 4c, are separated by approximately 0.6 nm. Upon propagation, these near-field lobes form the two well-separated lobes in the simulated far-field pattern, providing a model-based correspondence between the near-field separation and the observed far-field feature. Because coherent propagation can modify the lobe profiles and centroid positions, this correspondence is necessarily approximate. Taking the effective source-plane resolution of the microscopy, approximately 0.3 nm[17], as a conservative uncertainty in this spatial assignment, we estimate the characteristic near-field separation to be approximately $0.6 \pm 0.3$ nm. Hence, our simulations indicate that the observed far-field switching reflects a redistribution of the effective emission region on the subnanometre scale.

The present results show that the spatial origin of electron emission is not dictated solely by the local-field profile but can instead be governed by photon-energy-dependent selection among molecular emission channels with distinct spatial character. In the present system, this mechanism enables photon-energy-programmable control of the electron birth site at the subnanometric scale within the common local-field hotspot, thereby overcoming a central limitation of conventional local-field patterning. More broadly, this principle links the optical selection of excitation pathways to emission-site selection and opens a route to selective control in more complex molecular emitter systems. For example, in systems with multiple molecular electron emitters and spectrally distinguishable excitation pathways, it could enable selective control of the emission site of each emitter even within a common local-field region.

## Methods

### Experimental conditions and data acquisition

Electron emission patterns from single-molecule emitters formed at the apex of tungsten tips were recorded at room temperature under ultrahigh vacuum ($1 \times 10^{-10}$ mbar). The tungsten tip apex radius was typically on the order of 100 nm. To fabricate single-molecule electron emitters, we first deposited $C_{60}$ multilayers onto the tip. Subsequent application of a strong DC field removed most molecules and left only a small number of protrusions at the apex, from which single-molecule electron emitters formed stochastically[13].

The excitation laser operated at a repetition rate of 80 MHz. The excitation photon energy was tuned using an intracavity birefringent filter and then frequency-doubled in a β-barium borate (BBO) crystal, yielding pulses at 2.99 and 3.18 eV with an estimated duration of ~40 fs. The laser pulses were guided into the vacuum chamber and focused onto the tip apex. The focal spot size at the tip apex was measured to be ~2.1 μm ($1/e^2$ diameter) from the spatial profile of the laser-induced electron-emission signal as the tip position was scanned across the focus. These pulses induced electron emission from single-molecule emitters[14]. Excitation polarisation and intensity were controlled using a half-wave plate and a polariser; unless otherwise stated, the polarisation was aligned parallel to the tip axis. The polarisation angle $\theta_p$ is defined with respect to the tip axis.

For single-molecule electron emitters, the effective spatial resolution of the electron emission patterns in the present geometry, referred to the emitter, was estimated to be ~0.3 nm[17]. The higher photon energies used here (~3 eV) enabled operation at substantially lower optical powers than in our previous lower-photon-energy (~1.5 eV) experiments[14], typically around 1 mW. This reduced the thermal load and improved emitter stability, allowing measurements to be sustained for about 1 h before molecular desorption occurred. In the measurements presented here, the average number of detected electrons remained below 0.05 per pulse, so that multi-electron emission within a single pulse and space-charge effects were negligible. Further details of the experimental setup and sample preparation are provided in Sections S1 and S2 of the Supplementary Information.

**Image processing and analysis**

Emission images shown in Figs. 2–4 were displayed using a linear intensity scale clipped at the 97th percentile. Images were centred using the intensity centroid, computed after percentile clipping, and cropped to a fixed field of view. Contours shown in Figs. 2 and 3 were extracted from Gaussian-smoothed images ($\sigma$ = 3 pixels) using a fixed relative intensity threshold. The spatial configuration of the emission patterns was quantified using the normalised lobe separation, $d/d_0$, where $d$ is the distance between the centres of mass of the emission lobes and $d_0$ is the corresponding value for the leftmost image in each dataset, that is, the reference image for each measurement series.

**Orbital-resolved emission model and near-/far-field propagation**

The simulated emission patterns were obtained in three steps. For the first step, molecular orbitals of model molecular emitters under external DC fields were calculated using GAMESS[18]. The corresponding three-dimensional orbital wavefunctions were extracted from the GAMESS output using MacMolPlt[19]. For the second step, the observed emission pattern was modelled as an incoherent sum of orbital-resolved intensity contributions,

$I(r) \propto \sum_i w_i I_i(r),$

where $r$ denotes the position on the detector plane, $I_i(r)$ is the intensity pattern associated with orbital $i$, and $w_i$ are effective emission weights. Electrons excited in the metallic tip and transmitted through the surface barrier occupy a relatively narrow post-barrier energy distribution[6], represented in Fig. 4a by the asymmetric curves referred to as the energy-supply function. Only orbitals lying within this energy window contribute appreciably to emission. For each such orbital, the effective emission weight $w_i$ was defined as the product of the energy-supply factor evaluated at the corresponding orbital energy and a three-dimensional overlap integral between the molecular orbital and the evanescent continuation of the photoexcited metal-state wavefunction into the molecular region[15]. This evanescent metal-state wavefunction was modelled by an exponentially decaying form[20]. The starting position of this decay was determined from the field-modified surface potential barrier obtained from DFT calculations with VASP[21,22]. The decay constant $\kappa$ was evaluated at a representative photoexcited electron energy and treated as constant. Although $\kappa$ in principle depends on the normal energy component, tunnelling predominantly selects near-normal transmission, and a normal-incidence approximation was adopted.

For the third step, only the vacuum-side component of each orbital, extracted from the region outside a radius of approximately 1 nm from the molecular centre, was propagated using a propagator formalism for a uniform electric field[23,24]. This component was taken to represent the post-tunnelling wavefunction in vacuum. Momentum components opposite to the emission direction were removed before propagation, as they do not contribute to emission towards the detector. These two procedures and their influence on the calculated patterns are discussed further in Section S7 of the Supplementary Information. The tip-screen distance used in the simulation was set to 70 mm, consistent with the experimental distance of approximately 70 mm. The corresponding orbital-resolved far-field intensity patterns $I_i(r)$ were then obtained from the propagated wavefunctions, and the total far-field emission pattern was constructed as the incoherent weighted sum $\sum_i w_i I_i(r)$. Unless otherwise noted, the simulations shown in the main text were performed for a vertically stacked dimer in a two-atom-chain–two-atom-chain configuration with an azimuthal offset of 90°, under a DC field of 4.5 V nm$^{-1}$, using an energy-supply broadening of 200 meV. This parameter set was selected on the basis of comparison between simulated and experimental emission patterns across multiple physically plausible structural models and parameter values, for which it gave the closest overall agreement with both experimental pattern shapes and their 0.19 eV photon-energy separation. The molecular configurations considered, together with representative comparisons showing how the simulated patterns depend on molecular configuration and DC field, are described in Sections S4, S6, and S8 of the Supplementary Information. The propagation method provides stable and physically consistent results in the high-voltage regime, where electron trajectories are predominantly forward-directed, and has proven useful in earlier studies for reproducing characteristic far-field interference and emission features in related nano-tip systems[14,25]. At lower acceleration voltages, the propagated patterns became unstable in our calculations. Because the weak-field limit of the uniform-field propagator treatment is delicate and has been discussed critically in the literature[24], we restrict the present analysis to the 3000 V condition, under which the propagation remained stable and captured the essential symmetry and spatial characteristics of the emission. In this sense, the propagation approach is used to visualise the correspondence between near-field orbital structure and far-field emission patterns, rather than to provide a fully quantitative description of the complete tip geometry. Full computational details are provided in the Supplementary Information.

We also note that the simulated emission patterns in Fig. 2a exhibit slight asymmetry, although the nominal dimer geometry possesses twofold rotational ($C_2$) symmetry about the surface-normal, or dimer, axis. In the calculations, the molecular geometry was optimized without imposing point-group symmetry constraints before the electronic-structure calculation. The resulting small atomic displacements produce a slight deviation from exact $C_2$ symmetry, allowing closely spaced MOs with even and odd $C_2$ characters to mix. This mixing introduces weak asymmetry into the orbital densities and, consequently, into the simulated emission patterns. Further details of this symmetry breaking are provided in Section S9 of the Supplementary Information.

## References

1. Kubo, A., Onda, K., Petek, H., Sun, Z., Jung, Y. S. & Kim, H. K. Femtosecond imaging of surface plasmon dynamics in a nanostructured silver film. Nano Lett. 5, 1123–1127 (2005).
2. Dąbrowski, M., Dai, Y. & Petek, H. Ultrafast photoemission electron microscopy: Imaging plasmons in space and time. Chem. Rev. 120, 6247–6287 (2020).

3. Awada, C., Popescu, T., Douillard, L., Charra, F., Perron, A., Yockell-Lelièvre, H., Baudrion, A. L., Adam, P. M. & Bachelot, R. Selective excitation of plasmon resonances of single Au triangles by polarization-dependent light excitation. J. Phys. Chem. C 116, 14591–14598 (2012).
4. Krüger, M., Schenk, M. & Hommelhoff, P. Attosecond control of electrons emitted from a nanoscale metal tip. Nature 475, 78–81 (2011).
5. Rohmer, M., Bauer, M., Leißner, T. & Aeschlimann, M. Time-resolved photoelectron nano-spectroscopy of individual silver particles: Perspectives and limitations. Phys. Status Solidi B 247, 1132–1138 (2010).
6. Yanagisawa, H., Hengsberger, M., Leuenberger, D., Klöckner, M., Hafner, C., Greber, T. & Osterwalder, J. Energy distribution curves of ultrafast laser-induced field emission and their implications for electron dynamics. Phys. Rev. Lett. 107, 087601 (2011).
7. Aeschlimann, M., Bauer, M., Bayer, D., Brixner, T., García de Abajo, F. J., Pfeiffer, W., Rohmer, M., Spindler, C. & Steeb, F. Adaptive subwavelength control of nano-optical fields. Nature 446, 301–304 (2007).
8. Aeschlimann, M., Bauer, M., Bayer, D., Brixner, T., Cunovic, S., Dimler, F., Fischer, A., Pfeiffer, W., Rohmer, M., Schneider, C., Steeb, F., Strüber, C. & Voronine, D. V. Spatiotemporal control of nanooptical excitations. Proc. Natl Acad. Sci. USA 107, 5329–5333 (2010).
9. Yanagisawa, H., Hafner, C., Doná, P., Klöckner, M., Leuenberger, D., Greber, T., Hengsberger, M. & Osterwalder, J. Optical control of field-emission sites by femtosecond laser pulses. Phys. Rev. Lett. 103, 257603 (2009).
10. Kern, J., Großmann, S., Tarakina, N. V., Häckel, T., Emmerling, M., Kamp, M., Huang, J. S., Biagioni, P., Prangsma, J. C. & Hecht, B. Atomic-scale confinement of resonant optical fields. Nano Lett. 12, 5504–5509 (2012).
11. Baumberg, J. J., Aizpurua, J., Mikkelsen, M. H. & Smith, D. R. Extreme nanophotonics from ultrathin metallic gaps. Nat. Mater. 18, 668–678 (2019).
12. Urbieta, M., Barbry, M., Zhang, Y., Koval, P., Sánchez-Portal, D., Zabala, N. & Aizpurua, J. Atomic-scale lightning rod effect in plasmonic picocavities: A classical view to a quantum effect. ACS Nano 12, 585–595 (2018).
13. Yanagisawa, H., Bohn, M., Goschin, F., Seitsonen, A. P. & Kling, M. F. Field emission microscope for a single fullerene molecule. Sci. Rep. 12, 2714 (2022).
14. Yanagisawa, H., Bohn, M., Kitoh-Nishioka, H., Goschin, F. & Kling, M. F. Light-induced subnanometric modulation of a single-molecule electron source. Phys. Rev. Lett. 130, 106204 (2023).
15. Yanagisawa, H. Modulating single molecular electron sources with light: Opportunities and challenges. Appl. Phys. Lett. 126, 010501 (2025).

16. Gadzuk, J. W. Single-atom point source for electrons: Field-emission resonance tunnelling in scanning tunnelling microscopy. Phys. Rev. B 47, 12832–12839 (1993).
17. Rose, D. J. On the magnification and resolution of the field emission electron microscope. J. Appl. Phys. 27, 215–220 (1956).
18. Barca, G. M. J., Bertoni, C., Carrington, L., Datta, D., De Silva, N., Deustua, J. E., Fedorov, D. G., Gour, J. R., Gunina, A. O., Guidez, E., Harville, T., Irle, S., Ivanic, J., Kowalski, K., Leang, S. S., Li, H., Li, W., Lutz, J. J., Magoulas, I., Mato, J., Mironov, V., Nakata, H., Pham, B. Q., Piecuch, P., Poole, D., Pruitt, S. R., Rendell, A. P., Roskop, L. B., Ruedenberg, K., Sattasathuchana, T., Schmidt, M. W., Shen, J., Slipchenko, L., Sosonkina, M., Sundriyal, V., Tiwari, A., Galvez Vallejo, J. L., Westheimer, B., Włoch, M., Xu, P., Zahariev, F. & Gordon, M. S. Recent developments in the general atomic and molecular electronic structure system. J. Chem. Phys. 152, 154102 (2020).
19. Bode, B. M. & Gordon, M. S. MacMolPlt: A graphical user interface for GAMESS. J. Mol. Graph. Model. 16, 133–138 (1998).
20. Gadzuk, J. W. Anisotropic charge transfer rates in the scattering of oriented atoms or molecules from surfaces. Surf. Sci. 180, 225–236 (1987).
21. Kresse, G. & Hafner, J. Ab initio molecular dynamics for liquid metals. Phys. Rev. B 47, 558–561 (1993).
22. Kresse, G. & Furthmüller, J. Efficient iterative schemes for ab initio total-energy calculations using a plane-wave basis set. Phys. Rev. B 54, 11169–11186 (1996).
23. Lukes, T. & Somaratna, K. T. S. The exact propagator for an electron in a uniform electric field and its application to Stark effect calculations. J. Phys. C: Solid State Phys. 2, 586–592 (1969).
24. Whitcombe, A. R. Comments upon the exact propagator for an electron in a uniform electric field. J. Phys. C: Solid State Phys. 4, 1–2 (1971).
25. Yanagisawa, H., Ciappina, M., Hafner, C., Schoetz, J., Osterwalder, J. & Kling, M. F. Optical control of Young's type double-slit interferometers for laser-induced electron emission from a nano-tip. Sci. Rep. 7, 12661 (2017).

**Acknowledgements**

The DFT calculations were performed using computational resources at the Supercomputer Center, Institute for Solid State Physics, the University of Tokyo (2025-Ba-0066), and SQUID at the D3 Center, the University of Osaka. We thank Dr Paul Menczel for valuable discussions.

**Funding statement**

This work was supported by the Research Foundation for Opto-Science and Technology, the Nakatani Foundation, the Murata Science Foundation, JST PRESTO (Grant No. 1082208), and JSPS KAKENHI (Grant Nos. JP24H00816, JP24H00817, JP24H00818, JP24H00820, and JP25H00842). A part of this research is based on the Cooperative Research Project of Research Center for Biomedical Engineering. H.Y. acknowledges support from the Hamamatsu Campus 100th Anniversary Program at Shizuoka University.

**Author contributions**

H.Y. led the project. H.Y. conceived the project, designed the experiments, and developed the experimental methodology. H.Y. performed the measurements and analysed the data, with A.S. and P.B. contributing additional measurements. H.Y. developed the theoretical framework and performed the emission modelling, including pattern simulations and data analysis. R.K. and H.F. performed the VASP calculations. H.N.-K. performed the GAMESS calculations. H.Y., A.S., R.K., H.F., P.B., N.L., and H.N.-K. contributed to the interpretation of the results. H.Y. wrote the manuscript with input from all authors. All authors discussed the results and commented on the manuscript.

**Competing interests**

The authors declare no competing interests.

**Data availability**

Source data are provided with this paper. Additional data that support the findings of this study have been deposited in figshare and are available to the editors and reviewers during peer review. These data will be made publicly available upon publication.

**Code availability**

Custom code used for the simulations and data analysis in this study has been deposited in figshare and is available to the editors and reviewers during peer review. The code will be made publicly available upon publication.